\documentclass{camera}
\usepackage{graphicx}
\usepackage{epstopdf}
\usepackage{amsmath,amssymb}

\newcommand{\pt}           {\ensuremath{p_{\rm T}}}
\newcommand{\pp}           {pp}
\newcommand{\PbPb}         {\mbox{Pb--Pb}}
\newcommand{\pPb}          {\mbox{p--Pb}}

\newcommand{\Npart}        {\ensuremath{N_\mathrm{part}}}
\newcommand{\Nparttar}     {\ensuremath{N_\mathrm{part}^\mathrm{target}}}
\newcommand{\avNpart}      {\ensuremath{\langle N_\mathrm{part} \rangle}}

\newcommand{\Ncoll}        {\ensuremath{N_\mathrm{coll}}}

\newcommand{\qpa}          {\ensuremath{Q_{\rm pPb}}}

\newcommand{\Nch}          {\ensuremath{N_\mathrm{ch}}}

\begin{document}

%%%%%%%%%%%%%%%%%%%%%%%%%%%%%%%%%%%%%%%%%%%%%%%%%%%%%%%%
% The title, only the first letter capitalized; if you want to split it in
% two or more lines, put a \\ macro at each line break
% example: 
%   \title{Title: first line\\ second line}
%
\title{Particle production in \pPb\ collisions with ALICE at the LHC}

%%%%%%%%%%%%%%%%%%%%%%%%%%%%%%%%%%%%%%%%%%%%%%%%%%%%%%%%
% The author(s), separated by commas; do not put a
% comma before the last author, use instead the \and
% macro which produces a normal ``and'' in the
% caps/small caps context
%
\author{Alberica Toia, on behalf of the ALICE Collaboration}

%%%%%%%%%%%%%%%%%%%%%%%%%%%%%%%%%%%%%%%%%%%%%%%%%%%%%%%%
%
\organization{Goethe University Frankfurt and GSI}

\maketitle

\begin{abstract}
Measurements of the transverse momentum spectra of light flavor
particles at intermediate and high \pt\ are an important tool for QCD
studies.  In \pp\ collisions they provide a baseline for perturbative
QCD, while in \PbPb\ they are used to investigate the suppression
caused by the surrounding medium. In \pPb\ collisions, such
measurements provide a reference to disentangle final from initial
state effects and thus play an important role in the search for
signatures of the formation of a deconfined hot medium.  While the
comparison of the \pPb\ and \PbPb\ data indicates that initial state
effects do not play a role in the suppression of hadron production
observed at high \pt\ in heavy ion collisions, several measurements of
particle production in the low and intermediate \pt\ region indicate
the presence of collective effects.
\end{abstract}

%%%%%%%%%%%%%%%%%%%%%%%%%%%%%%%%%%%%%%%%%%%%%%%%%%%%%%%%
\section{Introduction}
The \pPb\ physics program has developed from crucial
control-experiment to study cold nuclear effects and to establish a
baseline for \PbPb\ to an area where to find groundbreaking
discoveries but also new challenges.

Nuclear modification factors measured by ALICE in minimum bias (MB)
\pPb\ collisions for charged particles~\cite{alice_RpA_new}, heavy
flavor and jets show no deviations from unity at high-\pt, demonstrating that the observed strong suppression in
\PbPb\ collisions is due to final state effects. 
However several
measurements~\cite{Abelev:2012ola,Abelev:2013bla,Abelev:2013haa,ABELEV:2013wsa}
of particle production in the low and intermediate \pt\ region 
can not be explained by an incoherent superposition of
\pp\ collisions, but rather call for coherent and collective effects.
As their strength increases with multiplicity, a more detailed
characterisation of the collision geometry is needed. Moreover, a knowledge of the geometry dependence is necessary to interpret the suppression pattern of charmonia (J/$\psi$ $\psi$(2S)) and  in \PbPb\ collisions relative to the effects observed in the nuclear medium produced in \pPb\ collisions.

\begin{figure}
\begin{center}
\includegraphics*[width=0.45\textwidth]{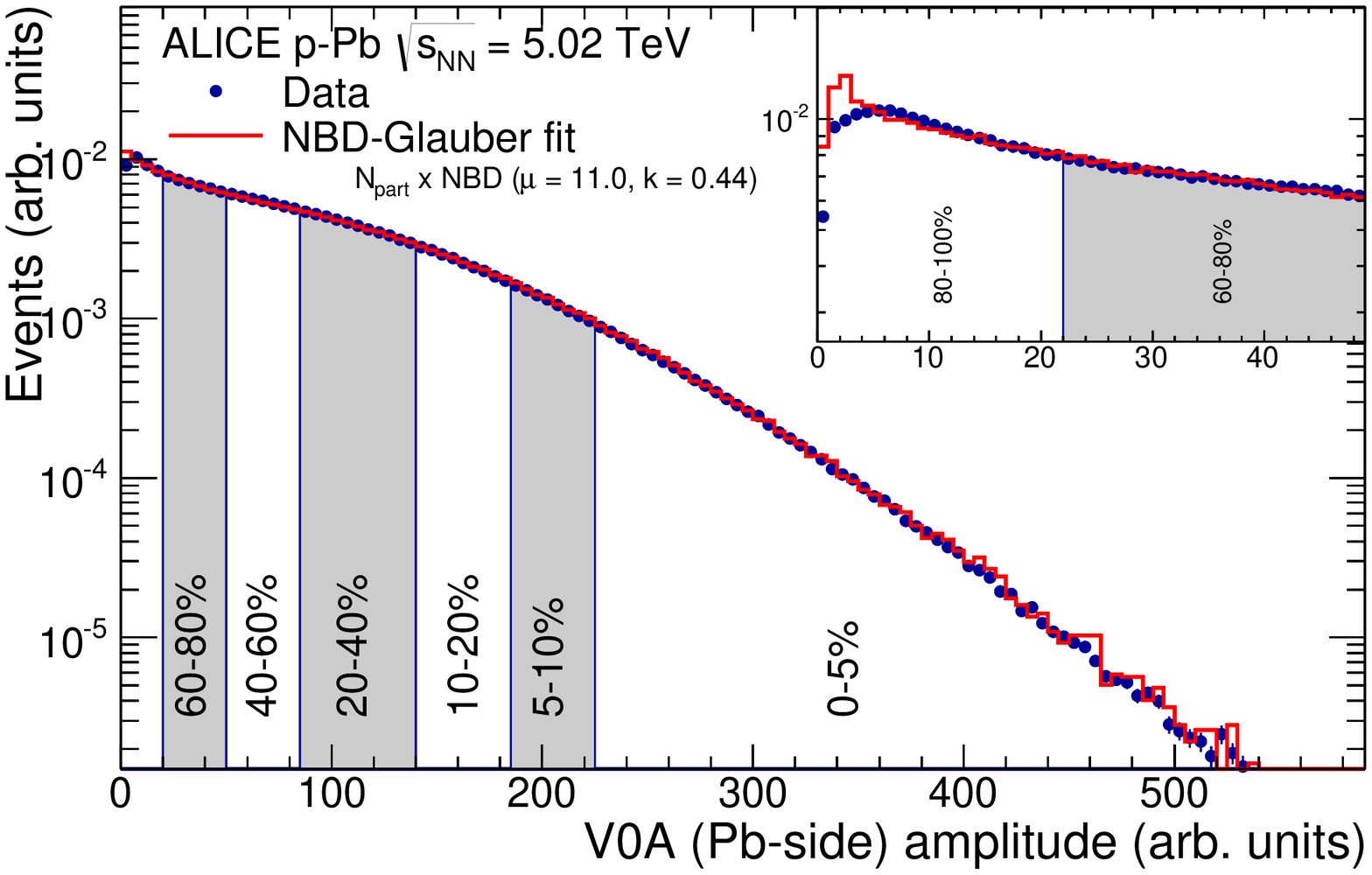}
\includegraphics*[width=0.49\textwidth]{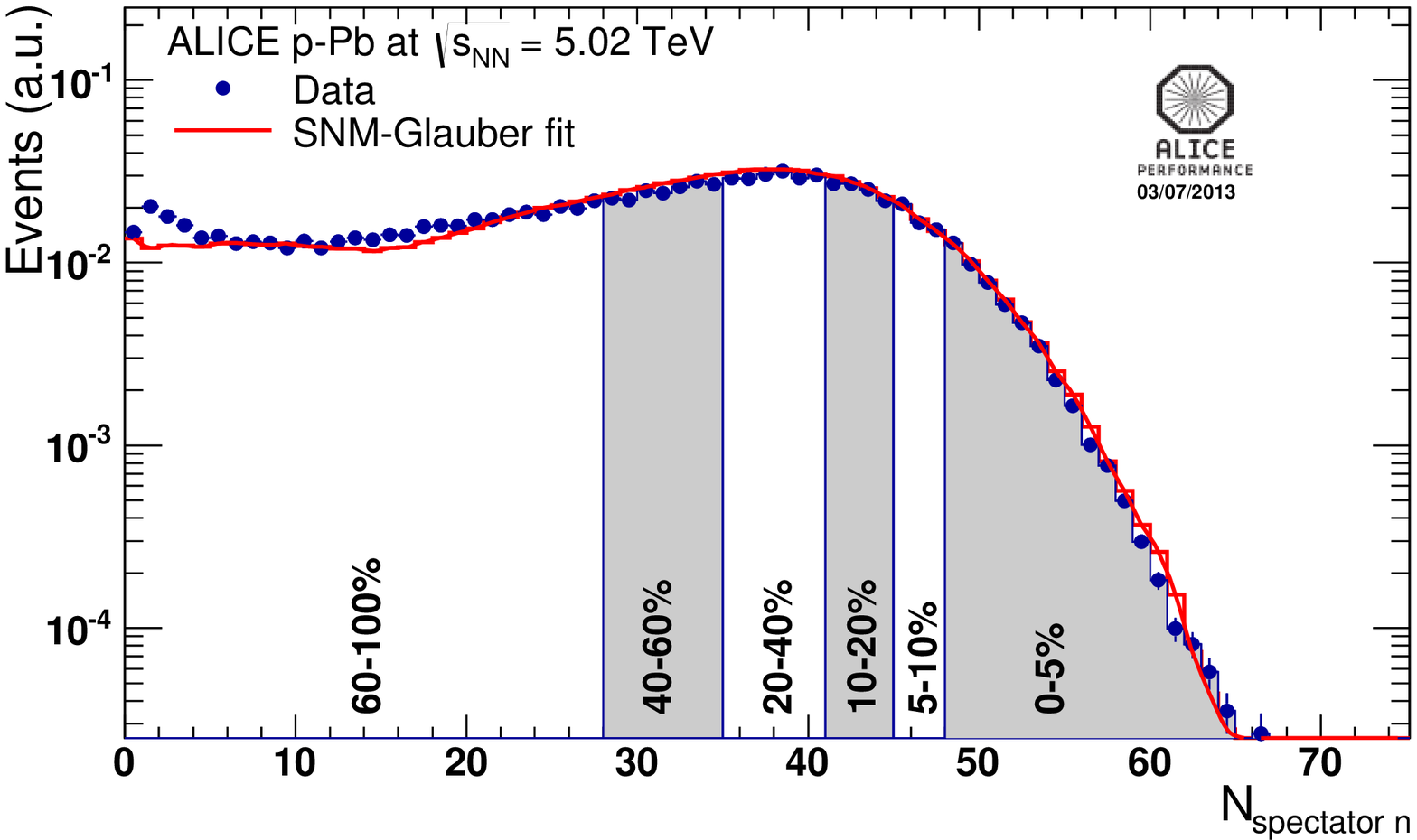}
\caption{Distribution of the sum of amplitudes in the V0 hodoscopes
  (left) and of the neutron energy spectrum measured in the ZN
  calorimeter (right) in the Pb-remnant side (A).  (Pb-going). Centrality
  classes are indicated by vertical lines.  The distributions are
  fitted with a Glauber model coupled to a NBD (left) or with the
  SNM-model (right) and are shown as a line.  }
\label{fig:glaunerNBDSNM}
\end{center}
\end{figure}

ALICE has carried out detailed studies of the centrality determination
in \pPb\ collisions, and the possible biases induced by the event
selection on the scaling of hard processes in a selected event sample.
The centrality determination consists in relating a Glauber model,
which calculates the geometric properties of the event (\Ncoll), to a
measured observable related to the event activity in a specific
rapidity region~\cite{Alice:Centrality}, via the conditional
probability to observe a certain activity for a given \Ncoll.
More specifically, particle production measured by detectors at
mid-rapidity can be modeled with a negative binomial distribution
(NBD). The zero-degree energy is related to the number of the slow nucleons emitted
in the nucleus fragmentation process, which we model with a Slow
Nucleon Model (SNM)~\cite{Alice:CentralityPA}.  
Fits of both models to our data are shown in Fig. \ref{fig:glaunerNBDSNM}.

However, the connection of the
measurement to the collision geometry has to be validated, eg by
correlating observables from kinematic regions causally disconnected
after the collision, or by comparing a Glauber MC with data for a known
process, as eg the deuteron dissociation probability at RHIC~\cite{Adare:2013nff}.

\begin{figure}[t!f]
 \centering
 \includegraphics[width=0.49\textwidth]{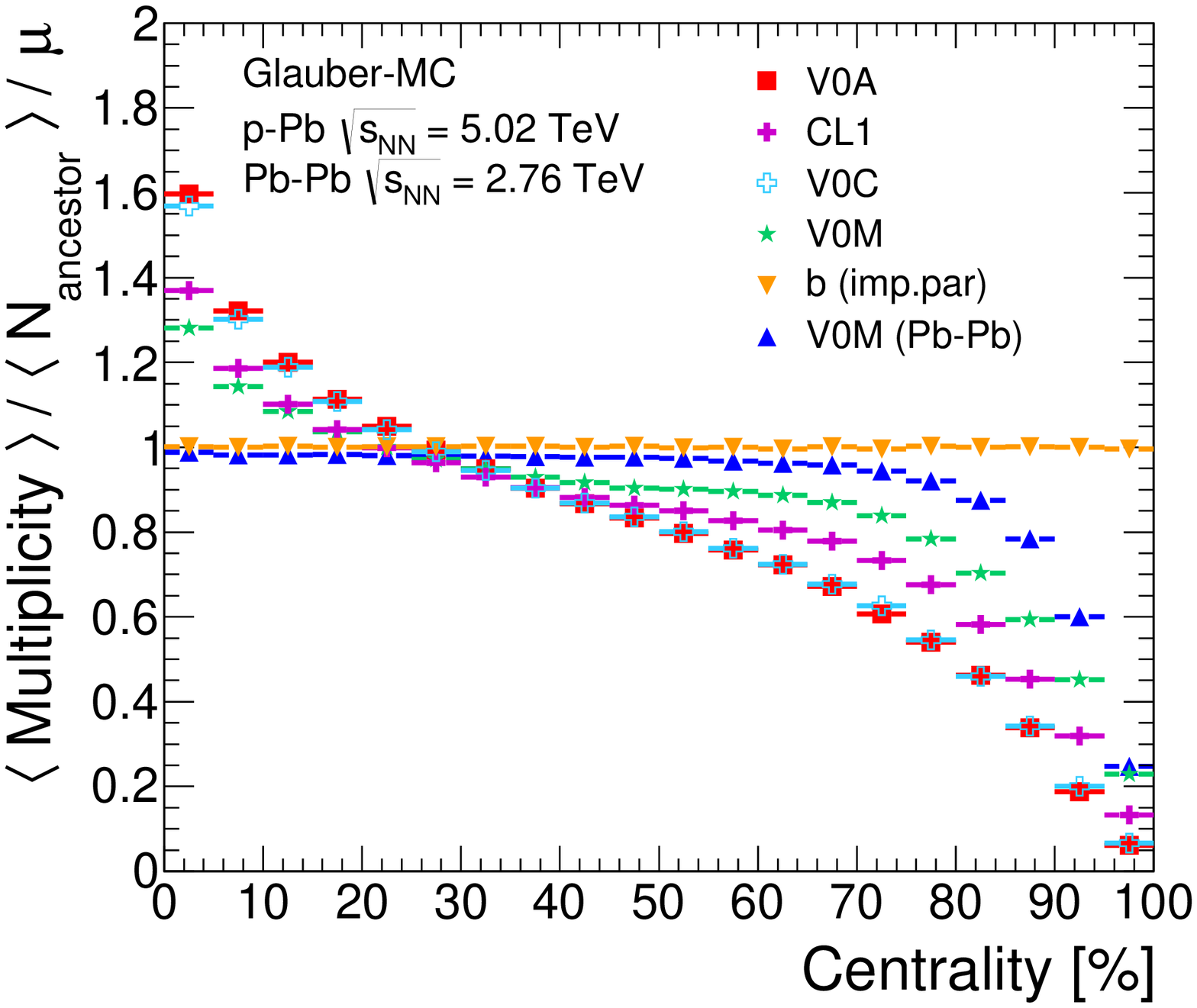}
 \includegraphics[width=0.49\textwidth]{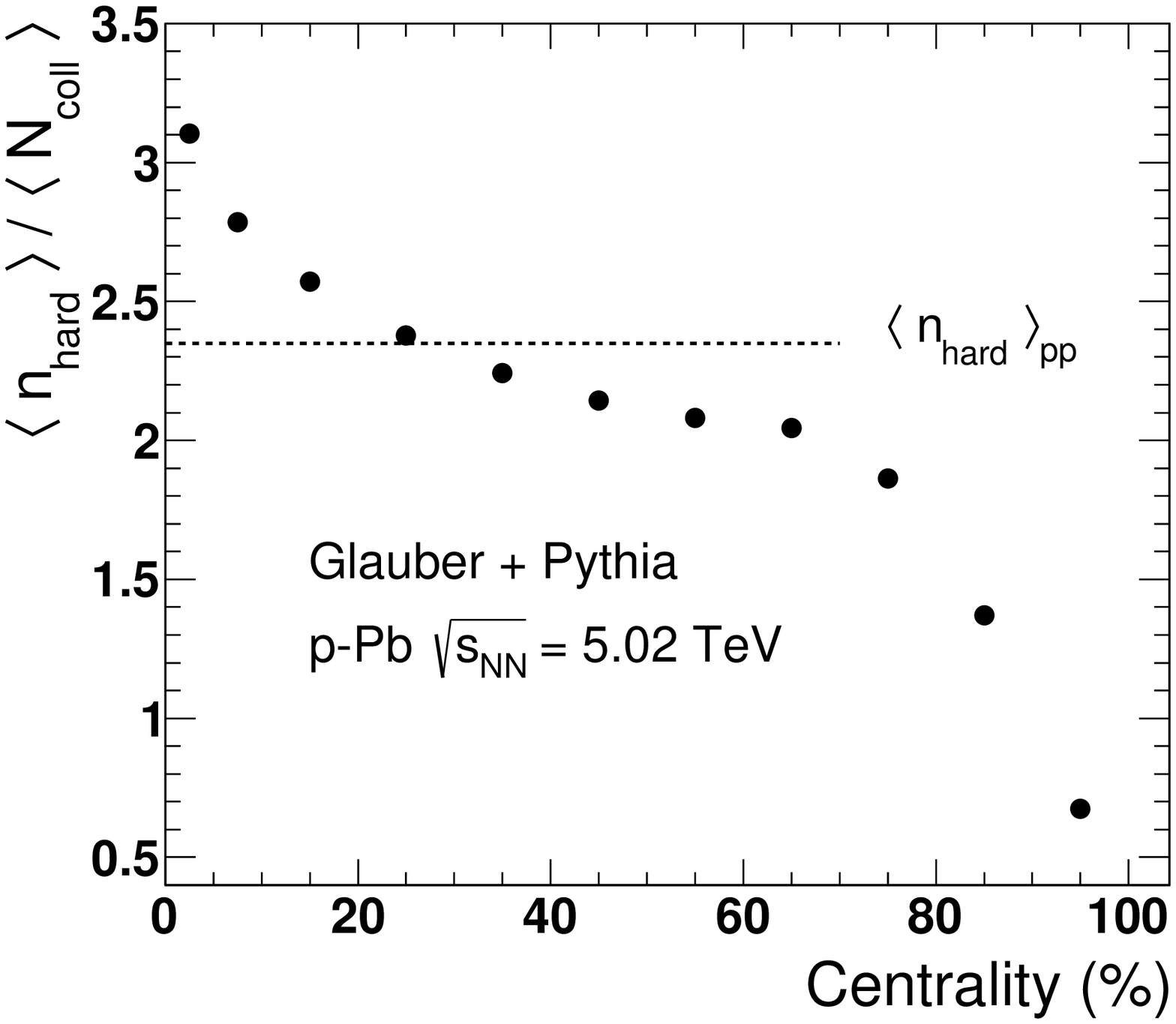}
 \caption{Left: Multiplicity fluctuation bias quantified as the mean
   multiplicity per $\avNpart / \mu$ from the NBD-Glauber MC in
   \pPb\ and \PbPb\ calculations. Right: Number of hard scatterings
   (MSTI(31) in PYTHIA6) per \Ncoll\ as a function of the centrality
   calculated with a toy MC that couples a pp PYTHIA6 calculation to a
   \pPb\ Glauber MC.
  \label{fig:multdist}}
\end{figure}

In addition the consistency of the approach must be demonstrated.  As
in general the selection in a system with large relative fluctuations
can induce a bias, one needs to identify the physics origin of the
bias in order to correct centrality dependent measurements.  In
\pPb\ collisions, the relative large size of the multiplicity
fluctuations has the consequence that a centrality selection based on
multiplicity may select a biased sample of nucleon-nucleon
collisions. In essence, by selecting high (low) multiplicity one
chooses not only large (small) average \Npart\, but also positive
(negative) multiplicity fluctuations per nucleon-nucleon collision.
This is shown on the left in Fig.~\ref{fig:multdist}.  These
fluctuations are partly related to qualitatively different types of
collisions, described in all recent Monte Carlo generators by impact
parameter dependence of the number of particle sources via
multi-parton interaction.  Hence, the biases on the multiplicity
discussed above correspond to a bias on the number of hard scatterings
($n_{\rm hard}$). As a consequence, for peripheral (central)
collisions we expect a lower (higher) than average number of hard
scatterings per binary collision, corresponding to a nuclear
modification factor less than one (greater than one), see
Fig.~\ref{fig:multdist}, right.  However also other types of biases
have an influence on the nuclear modification factor: the jet-veto
effect, due to the trivial correlation between the centrality
estimator and the presence of a high-\pt\ particles originating from
jets in the event; the geometric bias, resulting from the mean impact
parameter between nucleons rising for the most peripheral events
\cite{jia}.

\section{The ALICE approach}\label{sec:hybrid}  
The ALICE approach aims to a centrality
selection with minimal bias and, therefore, uses the signal in the Zero
Degree Calorimeter (ZNA). In this case we cannot establish a direct
connection using a well established model to the collision geometry but we can study the correlation
of two or more observables that are causally disconnected after the
collision, e.g because they are well separated in rapidity.

\begin{figure}
\begin{center}
\includegraphics*[width=0.495\textwidth]{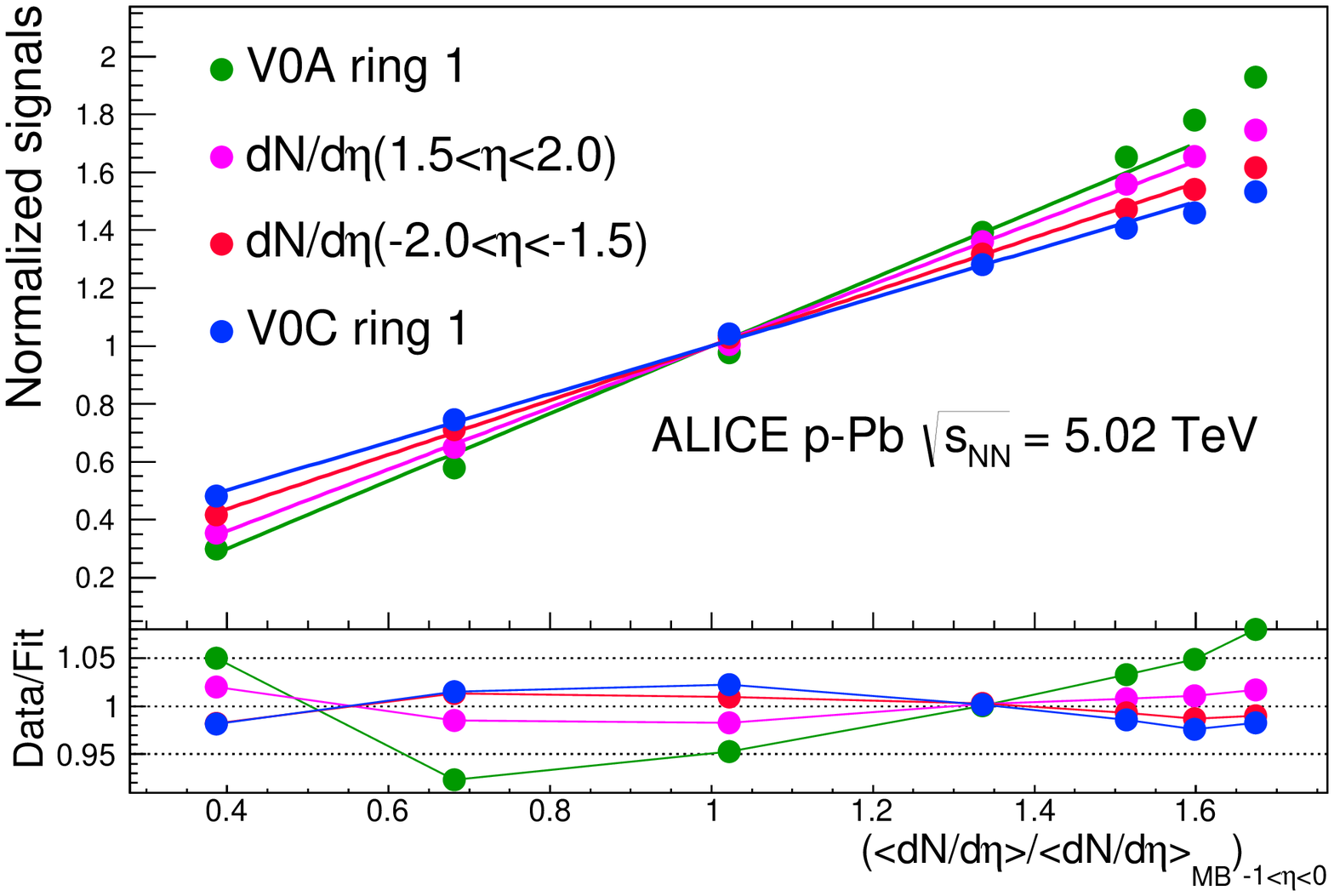}
\includegraphics*[width=0.35\textwidth]{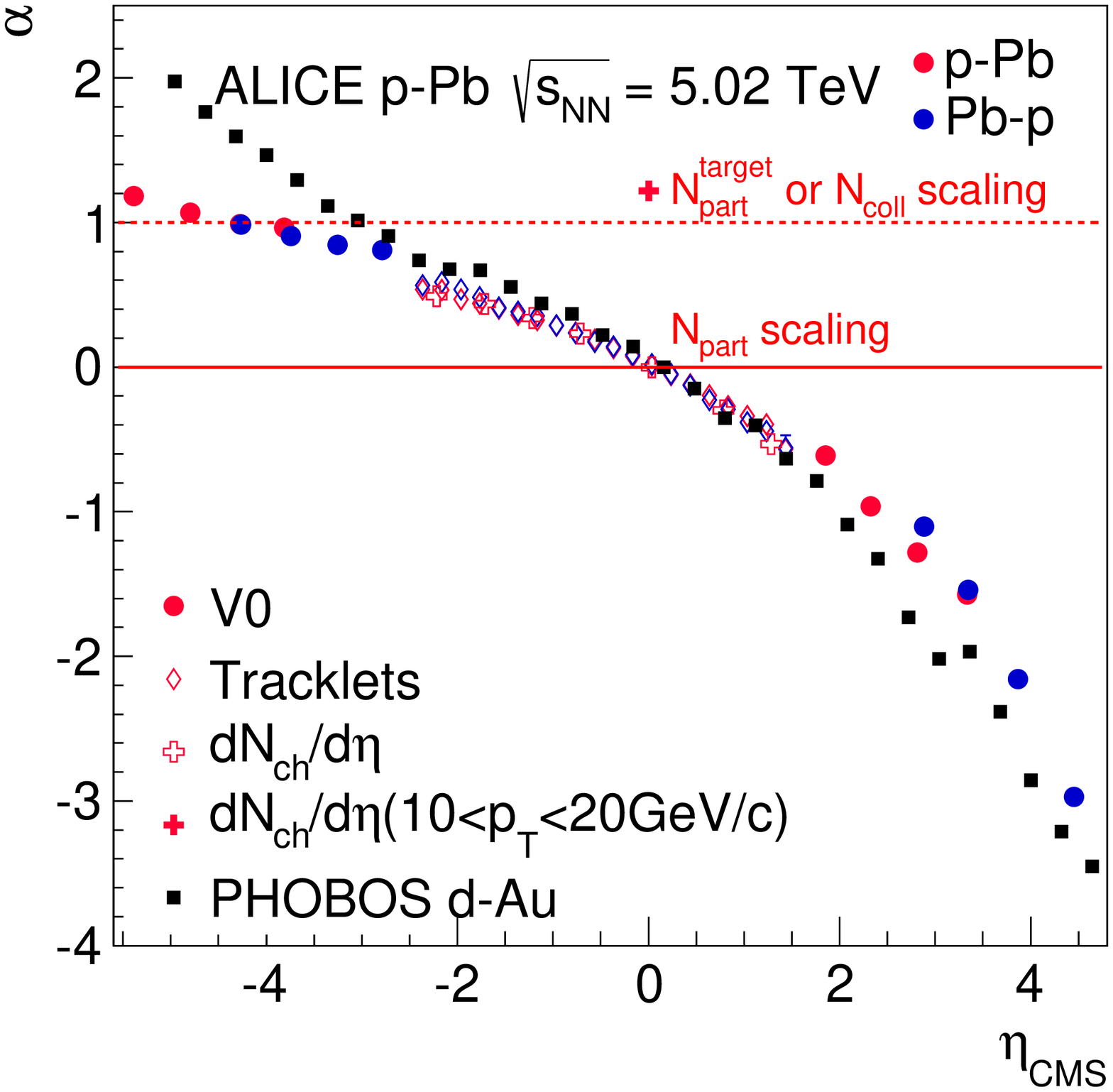}
\caption{Left: Normalized signal from various observables versus the
  normalized charged-particle density, fit with a linear function of
  \Npart. Right: Results from the fits as a function of the
  pseudorapidity covered by the various observables. The red
  horizontal lines indicate the ideal \Npart\ and \Ncoll\ geometrical
  scalings. The most central point is excluded from the fit, to avoid
  pile-up effects. PHOBOS points taken from \cite{Back:2004mr}.}
\label{fig:hybrid}
\end{center}
\end{figure}

In centrality
classes selected by ZNA, we study the dependence of various
observables in different $\eta$ and \pt\ regions on the charged
particle density at mid-rapidity. 
Fig.\ref{fig:hybrid} indicate a monotonic change of the
scaling with rapidity.  The correlation between the ZDC energy and
any variable in the central part shows unambiguously the connection of
these observables to geometry. 

Exploiting the findings from the correlation analysis described, we
make use of observables that are expected to scale as a linear
function of \Ncoll\ or \Npart, to calculate \Ncoll:
\begin{itemize}
\item $\Ncoll^{\rm mult}$: the multiplicity at mid-rapidity
proportional to the \Npart;
\item $\Ncoll^{\rm Pb-side}$: the target-going multiplicity
proportional to \Nparttar;
\item $\Ncoll^{\rm high-\pt}$: the yield of high-\pt\ particles at
mid-rapidity is proportional to \Ncoll.  
\end{itemize}
These scalings are used as an ansatz, in the ALICE socalled \emph{hybrid method} to calculate \Ncoll, rescaling
the MB value $\Ncoll^{\rm MB}$ by the ratio of the normalized signals
to the MB one.  We therefore obtain 3 sets of values of \Ncoll. The relative
difference does not exceed 10\%. 

\begin{figure}[t!f]
 \centering 
 \includegraphics[width=1.0\textwidth]{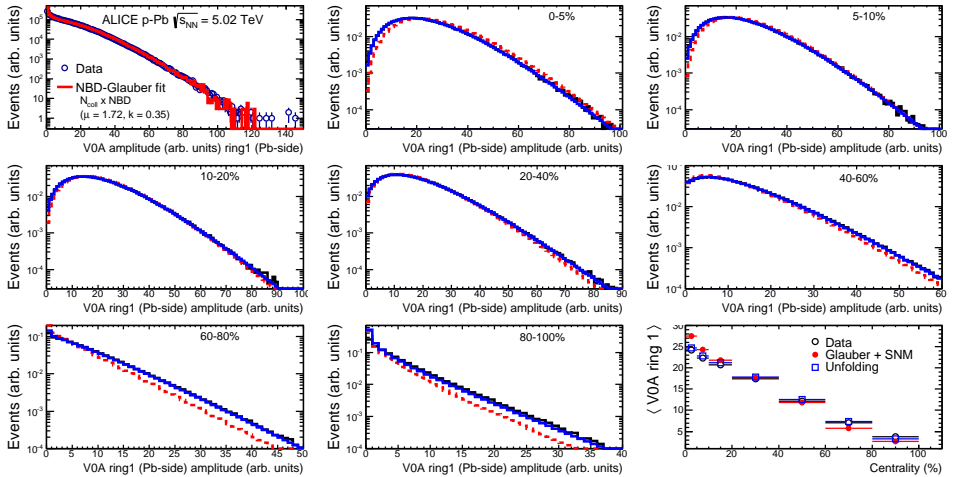} 
\caption{V0A ring1 signal distributions. 
The top left panel shows the distribution for MB events together with
a NBD-Glauber fit. The remaining panels show the distributions and
mean values for centrality classes selected with ZNA.  These are
compared to those obtained by the convolution of the ${\cal P}(\Ncoll
| cent_{\rm ZNA})$ distributions from the SNM with the NBD from the
NBD-Glauber fit to V0A ring1.  Data are also compared to the
distributions obtained with an unfolding procedure, where the \Ncoll\
distributions have been fitted to the data using the parameters from
the NBD-Glauber fit. The bottom right panel compares the mean values
of these distributions as a function of the centrality.
\label{fig:convolution}}
\end{figure}

We have performed a consistency check, correlating ZNA and V0A
centrality measurements, to establish their relation to the
centrality.  This is shown in Fig.~\ref{fig:convolution}.
The \Ncoll\ distributions for centrality classes selected
with ZNA, obtained from the SNM-Glauber fit, are convolved with the
NBD from the fit to the V0A distribution. These are compared to the
data and a good agreement is found.  Moreover, a good agreement is
also achieved with an unfolding procedure, where the
\Ncoll\ distributions have been fitted to the data using the
parameters from the NBD-Glauber fit.  For each V0A distribution
selected by ZNA, we find the \Ncoll\ distribution that, convolved with
the $\rm NBD_{\rm MB}$, fits the data, i.e. the parameters of the fit
are the relative contributions of each \Ncoll\ bin.  The existence of
\Ncoll\ distributions that folded with NBD agree with measured signal
distributions is a necessary condition for ZNA to behave as an
unbiased centrality selection. This procedure, which actually does not
work for a biased centrality selection (eg CL1) shows that the energy
measured by ZN is connected to the collision geometry.

\section{Physics Results}
\subsection{Nuclear Modification Factors}
\begin{figure}[t!f]
 \centering
\includegraphics[width=0.48\textwidth]{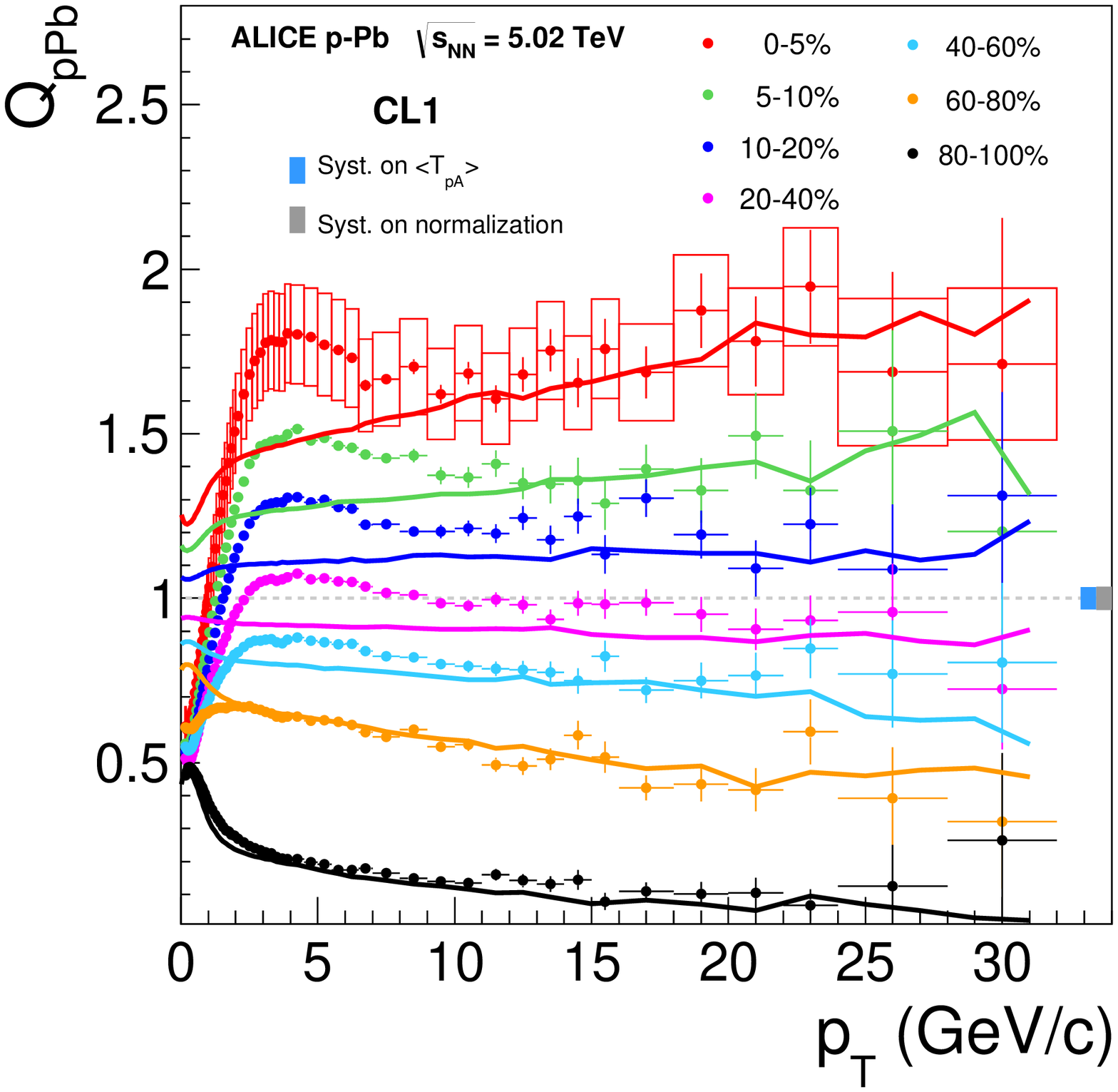}
\includegraphics[width=0.46\textwidth]{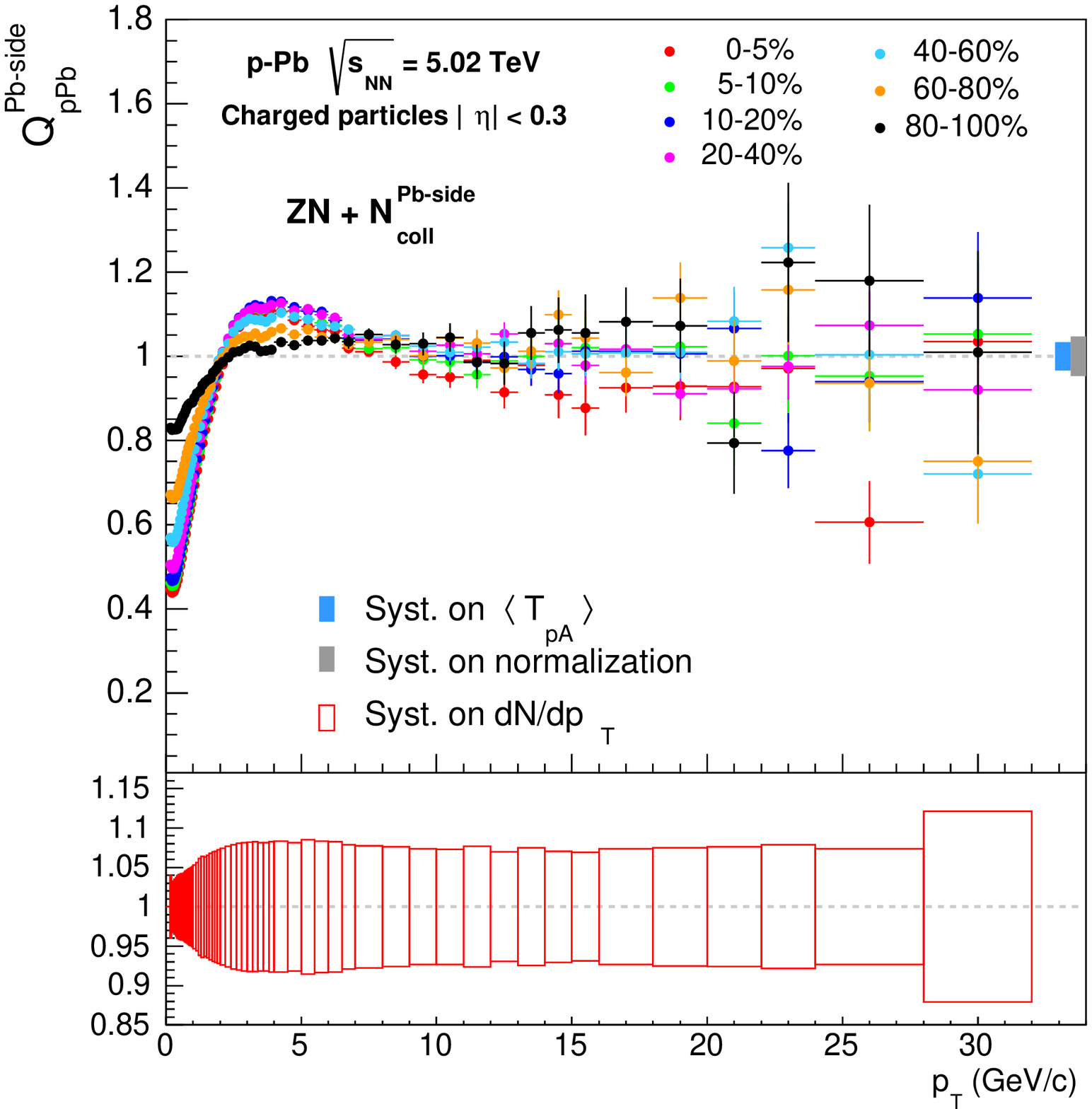}
 \caption{\qpa\ calculated with CL1 estimator (left), the lines are
   the G-PYTHIA calculations; with the hybrid method (right), spectra
   are measured in ZNA-classes and \Ncoll\ are obtained with the
   assumption that forward multiplicity is proportional to \Nparttar.
  \label{fig:QpAhybrid}}
\end{figure}

The nuclear modification factors \qpa\ calculated with
multiplicity based estimator (shown in Fig.\ref{fig:QpAhybrid} for
CL1, where centrality is based on the tracklets measured in
$|\eta|<1.4$) widely spreads between centrality classes. They also exhibit a
negative slope in \pt, mostly in peripheral events, due to the jet
veto bias, as jet contribution to particle production increases with \pt.  The \qpa\ compared
to G-PYTHIA, a toy MC which couples Pythia to a p-Pb Glauber MC, show
a good agreement, everywhere in 80-100\%, and in general at high-\pt,
demonstrating that the proper scaling for high-\pt\ particle
production is an incoherent superposition of \pp\ collisions, provided
that the bias introduced by the centrality selection is properly
taken into account, as eg in G-PYTHIA. 

With the hybrid method, using either the assumption on mid-rapidity
multiplicity proportional to \Npart, or forward multiplicity
proportional to \Nparttar, the \qpa\ shown in Fig.~\ref{fig:QpAhybrid},
are consistent with each other, and also consistent with unity for all
centrality classes, as observed for MB collisions, indicating the
absence of initial state effects. The observed Cronin enhancement is
stronger in central collisions and nearly absent in peripheral
collisions. The enhancement is also weaker at LHC energies compared to
RHIC energies.
%%%%%%%%%%%%%%%%%%%%%
%%%%%%%%%%%%%%%%%%%%%

\subsection{Charged particle density}
\begin{figure}[t!f]
 \centering 
 \includegraphics[width=0.57\textwidth]{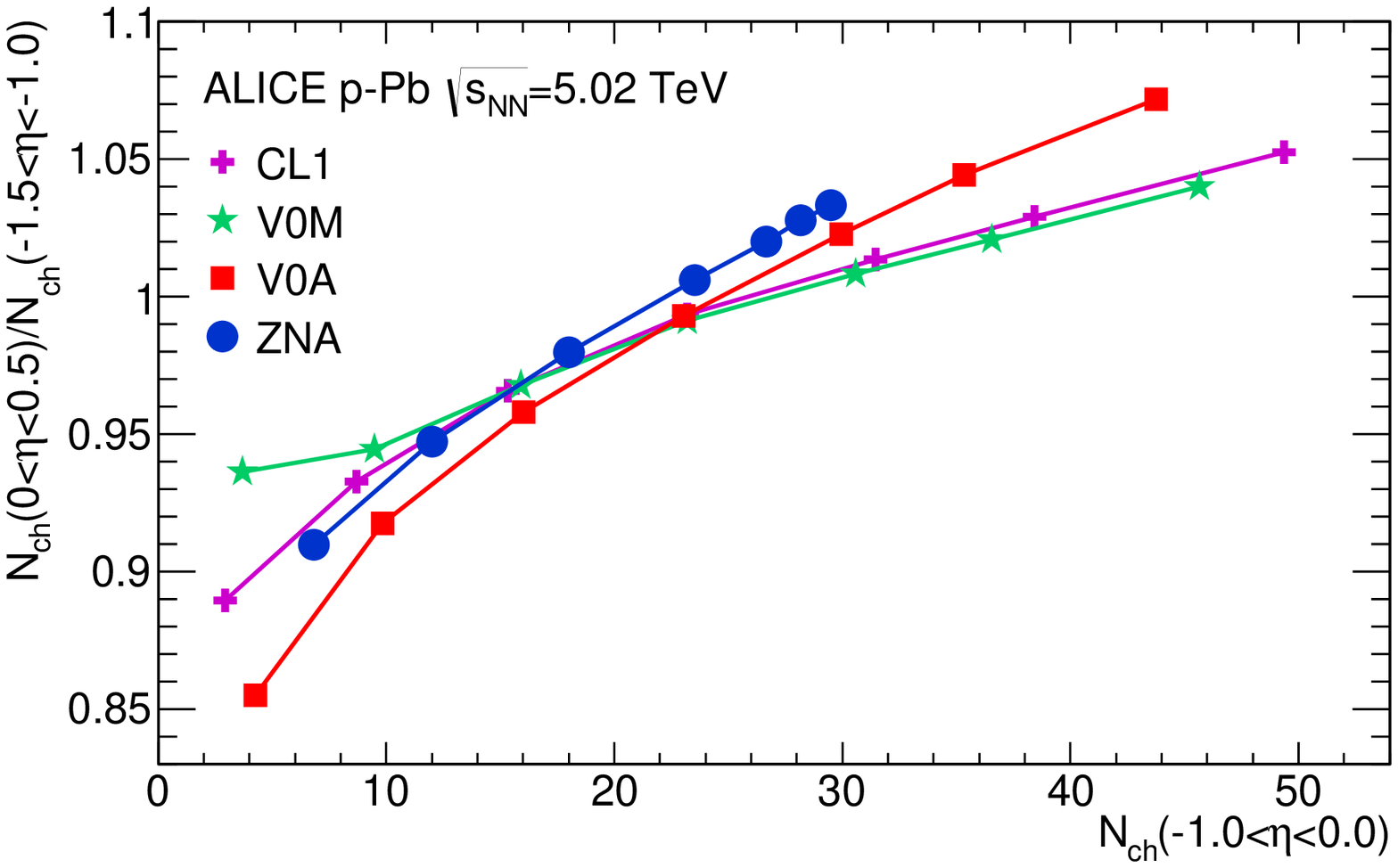}
 \includegraphics[width=0.4\textwidth]{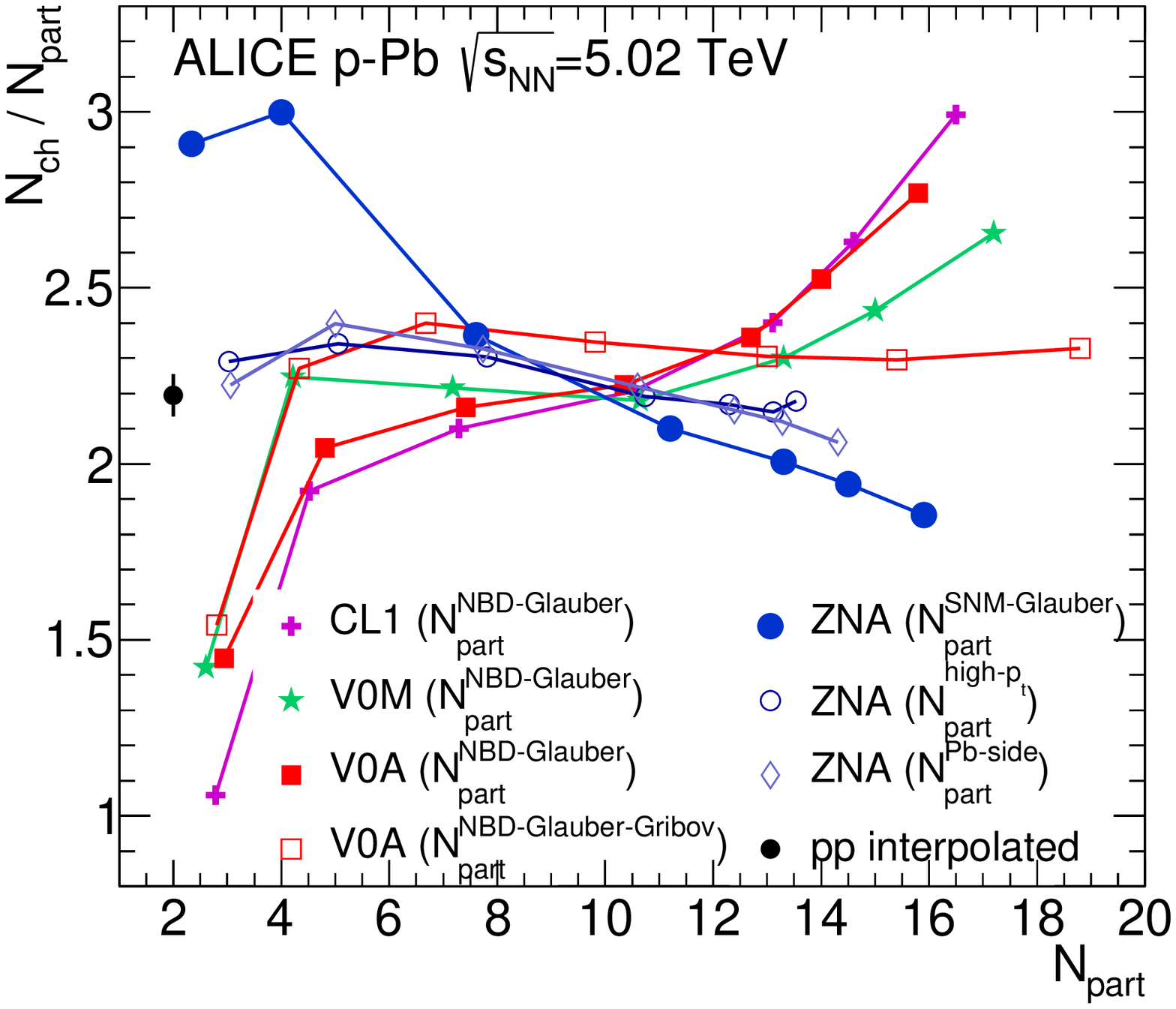}
 \caption{Left: Asymmetry of particle yield, as a function of the
   pseudorapidity density at mid-rapidity for various centrality
   classes and estimators. Right: Pseudorapidity density of charged
   particles at mid-rapidity per participant as a function of
   \Npart\ for various centrality estimators.
\label{fig:dndetaNpart2}}
\end{figure}

Charged particle density is also studied as a function of $\eta$, for
different centrality classes, with different estimators. In peripheral
collisions the shape of the distribution is almost fully symmetric and
resembles what is seen in proton-proton collisions, while in central
collisions it becomes progressively more asymmetric, with an
increasing excess of particles produced in the direction of the Pb
beam. We have quantified the evolution plotting the asymmetry between
the proton and lead peak regions, as a function of the yield around
the midrapidity (see Fig.~\ref{fig:dndetaNpart2}, left): the
increase of the asymmetry is different for the different estimators.
Fig.~\ref{fig:dndetaNpart2} right shows \Nch\ at mid-rapidity divided
by \Npart\ as a function of \Npart\ for various centrality
estimators. For Multiplicity-based estimators (CL1, V0M, V0A) the
charged particle density at mid rapidity increases more than linearly,
as a consequence of the strong multiplicity bias. This trend is absent
when \Npart\ is calculated with the Glauber-Gribov model, which shows
a relatively constant behavior, with the exception of the most
peripheral point.  For ZNA, there is a clear sign of saturation above
\Npart\ = 10, due to the saturation of forward neutron emission. None
of these curves points towards the \pp\ data point. In contrast, the
results obtained with the hybrid method, using either
\Nparttar-scaling at forward rapidity or \Ncoll-scaling for
high-\pt\ particles give very similar trends, and show a nearly
perfect scaling with \Npart, which naturally reaches the
\pp\ point. This indicates the sensitivity of the \Npart-scaling
behavior to the Glauber modeling, and the importance of the
fluctuations of the nucleon-nucleon collisions themselves.

\section{Conclusions}
Multiplicity Estimators induce a bias on the hardness of the pN
collisions.  When using them to calculate centrality-dependent \qpa,
one must include the full dynamical bias.  However, using the
centrality from the ZNA estimator and our assumptions on particle
scaling, an approximate independence of the multiplicity measured at
mid-rapitity on the number of participating nucleons is observed.
Furthermore, at high \pt\ the \pPb\ spectra are found to be consistent
with the pp spectra scaled by the number of binary nucleon--nucleon
collisions for all centrality classes. Our findings put strong
constraints on the description of particle production in high-energy
nuclear collisions.

%%%%%%%%%%%%%%%%%%%%%%%%%%%%%%%%%%%%%%%%%%%%%%%%%%%%%%%%
\end{document}